# Bio-inspired seismic metamaterials: Time domain simulations in transformed crystals

Ronald Aznavourian[1], Tania M. Puvirajesinghe[2,3], Stéphane Brûlé[4], Stefan Enoch[1] and Sébastien Guenneau[1,*]

[1]*Aix-Marseille Université, CNRS, Centrale Marseille, Institut Fresnel UMR 7249, 13013 Marseille, France*
[2]*Aix-Marseille Université, CRCM, Cell Polarity, Cell signaling and Cancer*
*"Equipe labellisée Ligue Contre le Cancer", Inserm, U1068, Marseille, F-13009, France*
[3]*Institut Paoli-Calmettes, F-13009, France*
[4] *Dynamic Soil Laboratory, Ménard, 91620 Nozay, France*
*\*sebastien.guenneau@fresnel.fr*

**Abstract:** We introduce the concept of transformation crystallography which consists of the application of geometric transforms to periodic structures. We consider motifs with three-fold, four-fold and six-fold symmetries according to the crystallographic restriction theorem. Furthermore, we define motifs with five-fold symmetry such as quasi-crystals generated by a cut-and-projection method. We analyze elastic wave propagation in the transformed crystals and (Penrose-type) quasi-crystals with the finite difference time domain freeware SimSonic. We consider geometric transforms underpinning the design of seismic cloaks with square, circular, elliptical and peanut shapes in the context of triangular, square and honeycomb crystals. Interestingly, the use of morphing techniques leads to the design of cloaks with interpolated geometries reminiscent of Victor Vasarely's artwork. Employing the case of transformed graphene-like (honeycomb) structures allows one to draw useful analogies between large-scale seismic metamaterials such as soils structured with columns of concrete or grout with soil and nanoscale biochemical metamaterials. We further point out similarities between cloaks for elastodynamic and hydrodynamic waves and cloaks for diffusion (heat or mass) diffusion processes, notably with respect to invisibility and protection. Experimental data extracted from field test analysis of soil structured with boreholes demonstrates the application of bio-inspired seismic metamaterials. We conclude that these novel materials hold strong applications in biophysics and geophysics.

**1. Introduction**

In 2011, the Israeli physicist Dan Shechtman received the Nobel Prize in chemistry for his discovery of a phase of an aluminum-manganese alloy with a five-fold symmetry originally carried out in 1984. In this breakthrough article[1] Shechtman together with his colleagues Blech, Cahn and Gratias discovered that this crystalline-like substance goes beyond the crystallographic restriction theorem that states that the only rotational symmetries allows for a crystalline pattern (i.e. a discrete system of points which has a translational symmetry) display two-fold, three-fold, four-fold and six-fold symmetries. One year previous to Shechtman's Nobel Prize, the Dutch-British physicist Sir Andrei Konstantinovich Geim and his Russian-British colleague Sir Konstantin Sergeevich Novolesov received the Nobel Prize in physics for their discovery of graphene in 2004[2]. Graphene is a one-atom thick layer of graphite (a crystalline form of carbon) with six-fold symmetry. A top view of graphene shows a honeycomb lattice as shown in Figure 1(a), consisting of one atom of carbon at each vertex of the 6-ring structure. Geim and Novoselov demonstrated the difference between graphene and graphite by using adhesive to isolate graphene (less than one nanometer thick) sheets away from graphite. Achieving single layers of graphene typically requires multiple exfoliation steps. Importantly, the acoustic and thermal properties of graphite, and thus graphene, are highly anisotropic, since phonons propagate quickly along the tightly-bound planes, but are slower to travel from one plane to another. Anisotropy is actually the essence of metamaterials, which make possible control of wave trajectories thanks to resonant subwavelengths elements (typically a few tenths of nanometers in size for visible light), enabling extreme effective tunable anisotropy.

The discoveries of quasi-crystals in 1984 and graphene in 2004 have revolutionized material sciences and given birth to new research areas in chemistry, biosciences, mechanical sciences and optics, amongst many disciplines. Interestingly, graphene flakes, which can be as small as a few tenths of nanometres in size, self-organize like clay in stratified soils, as noted in[3], and allow enhanced control of drug diffusion through effective anisotropic conductivity. Recently characterized seismic metamaterials have demonstrated a unique ability to control propagation of surface Rayleigh waves in soils structured on a metre scale[4], also via dynamic anisotropy. Analogies of models of diffusion in graphene oxide flakes and clay proposed in[3], suggest models of bio-inspired seismic metamaterials, which is the main focus of the present topical article. The contents of the article are as follows: we firstly introduce the concept of transformation crystallography and quasi-crystallography, with examples of transformed lattices with two-fold, three-fold, four-fold symmetries, and an example of transformed pentagon tilting in the plane. We then numerically investigate elastodynamic wave propagation with finite-difference-time-domain (FDTD) software SIMSONIC in large scale structured soils deduced from transformation crystallography. Our graphene-like seismic cloaks are also studied using a morphing algorithm [5]. We then consider similarities between invisibility cloaks for mechanical waves (counterpart of Pendry's electromagnetic cloak [6]) and thermal diffusion processes that have been experimentally validated in Marseille by our research group at Institut Fresnel and in Karlsruhe (research group of Martin Wegener at KIT). We further use some transmission electron microscopy images of flakes of a modified form of graphene (so-called graphene oxide, fabricated and characterized within the research group of Tania Puvirajesinghe at Institut Paoli-Calmettes in Marseille, for application in cancer therapy [3]), as a basis for numerical simulations of propagation and localization of elastodynamic waves within a scaled up version (by a factor of one million) of graphene flakes that can be considered as a bio-inspired seismic metamaterial. We conclude our paper with a review of field experiments carried out by the engineering group, Menard, with results obtained and analyzed by the soil dynamic laboratory of Stéphane Brûlé in 2012. All throughout this article, we point out analogies that can be drawn between crystallography, biology, wave physics and geophysics.

## 2. Transformation crystallography and quasi-crystallography

In this section, we introduce new concepts of transformation crystallography and quasi-crystallography. Before we do this, it is important to recall some counterintuitive properties of the latter. A remarkable feature of the quasi-crystalline tiling patterns is that their assembly is obligatory non-local. Namely, in assembling the patterns, it is necessary, in certain examples, to examine the state of the pattern multiple atoms away from the point of assembly, if one is to ensure to not commit a serious error when assembling the components together. In nature, a crystalline configuration is one that consists of the lowest energy status. With quasicrystal growth, the state of lowest energy is much more difficult to find, as compared with periodic structures, and the best arrangement of the atoms cannot be discovered by simply adding atoms sequentially with the assumption that each individual atom can solve its own minimizing problem. Instead, a global problem has to be solved. Such tilings were previously studied by Roger Penrose, who found many interesting mathematical properties related, as we will see, to arithmetic and logic. It should be noted that quasicrystalline substances also exhibit symmetry in three-dimensions, and not only in the plane, giving a forbidden icosahedral symmetry (these analogues of the Penrose tilings had been found by Robert Amman in 1975). Numerous mathematicians spent time on hobbies, pastimes and everyday activities, which often led to discoveries. Amongst them, the great number theorist Russel whose mathematical works inspired the world renowned books and puzzles of Lewis Carroll. In the early seventies, Roger Penrose was considering the question of covering the Euclidean plane with polygonal shapes, where there is a given finite number of different such shapes.

The initial task was to determine whether to cover the plane completely or not, without gaps or overlaps, using just these shapes and no others. It is well-known that such tilings are possible using just squares, or just equilateral triangles, or just regular hexagons, but not using just regular pentagons, as was described above. Many other single shapes will tile the plane, such as irregular pentagons, but with a pair of shapes the tilings can become more elaborate. It is worth noting that the same 'versatile' shape tile both periodically and non-periodically the Euclidean plane, is a property shared by many other single tile shapes and sets of tile shapes. The question remains do single or sets of tiles tile the plane in a non-periodically structure. In 1971, the American mathematician Raphael Robinson exhibited a tiling with six shapes, which tile the plane only in a non-periodic fashion. His works were inspired from those of the Chinese-American logician Hao Wang who addressed the question of whether or not there is a decision procedure for the tiling problem, that is to say, is there an algorithm for deciding whether or not a given finite set of different polygonal shapes will tile the entire plane.

Hao Wang was able to show that there indeed would be such a decision procedure if it could be shown that every finite set of distinct tiles used to tile a plane in a particular manner, will in fact also tile the plane periodically. In 1966, following directions and suggestions from Hao Wang had suggested, Robert Berger was able to show that there is in fact no decision procedure for the tiling problem: the tiling problem is a part of what is called non-recursive mathematics. Thus, it follows from Hao Wang's earlier result that an aperiodic set of tiles must exist. The example validating this was exhibited by Berger and involved originally 20426 different tiles. Berger was then able to reduce this number to 104 tiles in 1964. In 1971, we have seen that Raphael Robinson reduced this number down to six. Finally, Roger Penrose found in 1974 an aperiodic tile made of two tiles.

The question of whether or not it is possible to tile the Euclidean plane aperiodically with a single tile remains an open question. According to Roger Penrose, one method of achieving this would be to assign the vertices of the tile as points in the complex plane, and these points may be given as algebraic numbers. Indeed, N. G. de Bruijn showed that such patterns could be built thanks to a cut and projection method that was developed, among others, by M. Duneau and A. Katz. Let us briefly recall this cut and projection method which is not an isomorphism. A strip is obtained by shifting a unit square $Y^2 = ]0;1[^2$ along $E\|$. Then, for almost all positions of the line $E\|$, this strip contains a unique broken line (made of edges of the lattice), which joins exactly all the vertices inside the strip. The orthogonal projection $P\|$ of this line on $E\|$ gives a tiling, the two tiles being the projections of the vertical and horizontal edges of the unit square $Y^2$. Consider now the slope of the line $E\|$ with respect to the canonical basis of $R^2$. It should be checked if the tiling is periodic and if the slope is a rational number. An important property of such patterns is the local isomorphism property, introduced by M. Duneau and A. Katz as a property of Penrose tilings, which asserts that any finite patch of tiles that belongs to a tiling appears infinitely many times in any tiling defined through a strip with the same irrational slope. To prove this property, one has to consider the line $E\perp$, orthogonal to $E\|$. A finite patch of tiles in a given tiling is the projection of a finite broken line, the projection of which on $E\perp$ is strictly smaller than the projection of the whole strip. Then, there exists a non-empty open set of translations in $E\perp$ that keep the projection of the finite broken line inside the strip. Because the orthogonal projection $P\perp$ of $Z^2$ on $E\perp$ is dense, there are infinitely many translations of $Z^2$ which map the finite broken line inside the strip, which give infinitely many copies of the initial patch of tiles. In the same way, one can show that any finite patch of tiles that appears in a tiling appears in all tilings with the same slope.

There are interesting questions concerning the mean distance between copies of a given patch. The density of the copies of a patch depends on arithmetical properties of the slope. Indeed, it can be seen that when the slope of $E\|$ is given by an algebraic number, the mean distance between two copies is proportional to the size of the patch. On the contrary, when the slope is given by a Liouville number (non-algebraic irrational number), the distance can quickly grow arbitrarily. It should be noted that if the unit square is replaced by any other unit cell of the crystal lattice $Z^2$ (say $KY^2$, where K is a positive integer), new tilings involving the projections of the two edges of this cell are obtained. This is closely related to self-similarity properties of the Penrose tilings i.e. each fragmented geometric shape can be subdivided in parts, each of which is (at least approximately) a reduced-size copy of the whole. However, the projection method should not be restricted to these tilings since more general strips also generate tilings involving a finite number of tiles.

Furthermore, one can easily generalize this method to higher dimensions. To get a quasi-periodic pattern in $R^n$ issued from a projection of a periodic pattern in $R^m$ (m > n), one has to consider the subspace $E\| = R^n$ and $E\perp$ its orthogonal subspace in $R^m$. The Penrose tilings are issued from a cut and projection of $R^4$ (or $R^5$) in $R^2$ and their three-dimensional equivalents (the icosahedral structures) from a cut and projection of $R^6$ (or $R^{12}$) in $R^3$. These strange properties are related to abstract mathematical objects introduced by Benoit Mandelbrot: let us first consider the mapping $f(z)=z^2+c$ where z and c are two complex numbers. It is easy to see that for certain choices of the given complex number c, the mapping remains bounded (to be precise this set was found by Julia and Mandelbrot considered the mapping $f(z)=z^2+c$). Such a set of values of c defines what is called the Mandelbrot set.

    Now that we have set the scene of periodic and aperiodic tilings of the plane and their links with geometric transformations (cut-and-projection and for quasi-crystals and recursive mappings of the complex plane for fractals), we can turn our attention to transformation

crystallography and quasi-crystallography. Firstly, we recall the geometric transform for elliptical cloaks[9]:

$$f:(r,\theta) \longrightarrow (r',\theta') = (\alpha r + \beta, \theta), \text{ where } \forall 0 < r < R_2 = \sqrt{a_2^2 \cos^2\theta + b_2^2 \sin^2\theta}$$

$$\alpha = (R_2 - \beta)/R_2, \text{ with } \beta = \sqrt{a_1^2 \cos^2\theta + b_1^2 \sin^2\theta},$$

which maps the area within the ellipsis of eccentricities $a_2$, $b_2$ onto an elliptical corona delimited by ellipses of eccentricities $a_1$, $b_1$ and $a_2$, $b_2$.

Applying this geometric transform to the periodic structure shown in Figures 1(a), we obtain Figure 1(b) when $a_1=b_1$ and $a_2=b_2$ and Figure 1(c) when $2a_1=b_1$ and $2a_2=b_2$. Figures 1(d) and (e) are obtained by making r angle dependent.

Invisibility cloaks of arbitrary shapes described by Fourier series have been proposed in[10]:

$$g:(r,\theta) \longrightarrow (r',\theta') = (\alpha r + \beta, \theta), \text{ where } \forall 0 < r < R_2(\theta) = \sum_{k=1}^{n} a_2^k \cos(k\theta) + b_2^k \sin(k\theta)$$

$$\alpha = (R_2 - \beta)/R_2, \text{ with } \beta = \sum_{k=1}^{n} a_1^k \cos(k\theta) + b_1^k \sin(k\theta),$$

which maps the area within the closed curve of $R_2(\theta)$ onto an corona delimited by closed curves $R_1(\theta)$ and $R_2(\theta)$.

One can then deduce the metric tensor $T = J^T J / \det(J)$ in the transformed coordinates from the computation of the Jacobian matrix $J(r',\theta') = \partial(r,\theta) / \partial(r',\theta')$. This is also well-known in transform optics.

However, here we apply these transforms to crystalline and quasi-crystalline structures in order to generate a new class of transformed crystals and quasi-crystals.

In Figures 1 to 5 we demonstrate the application of geometric transforms f and g to honeycomb (Figure 1), square (Figure 2), triangular (Figure 3) and pentamode-like (Figure 4) lattices.

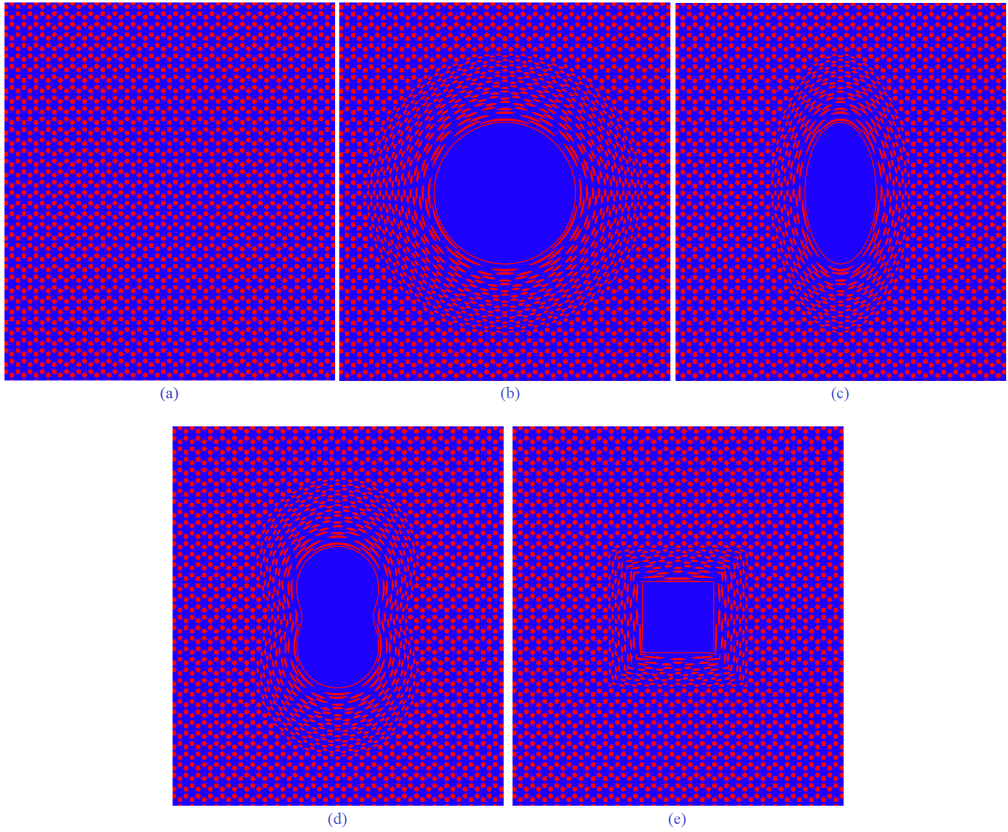

Figure 1: Graphene-like honeycomb lattice and transformed honeycomb lattices: (a) honeycomb lattice; (b-e): Transformed honeycomb lattice with a circular cloak (b), an elliptic cloak (c), a peanut cloak (d), a square cloak (e).

We obtain a collection of transformed lattices reminiscent of paintings by Victor Vasarely[11], an artist renown for his artwork on optical illusions. It could be considered that Vasarely initiated transformation crystallography in an art form at approximately the same period as Victor Veselago introduced negative refraction[12]. Although the painter performed his geometric transforms with a pen and a piece of paper, the result of our computations appears to be strikingly similar to his paintings. It is important to precise at this point that other parallels have been drawn in[13] between complementary media and Vasarely's paintings. Therein, the crystallographic restriction theorem was invoked to keep a balance between overall positive and negative refractive index materials as its application implies that only checkerboards with either rectangular, square or (equilateral) triangular cells can lead to perfect imaging devices associated with overall zero optical path-length. Although the authors of [13] did not realize it at that time, they were essentially making use of transformational crystallography tools i.e. coordinate changes in periodic structures (in that case periodic sign-shifting checkerboards). An important feature that unfortunately falls beyond the scope of the present article is the application of the Floquet-Bloch theorem in transformed lattices. A Floquet-Bloch wave function u would be of the form $u(f(\mathbf{x}+\mathbf{T}))=u(\mathbf{x}) \exp (i f(\mathbf{k}.\mathbf{T}))$, where $i^2=-1$, $\mathbf{x}$ is the position vector and $\mathbf{k}$ and $\mathbf{T}$ are respectively the Bloch and lattice vectors in the periodic reciprocal space and f is the mapping onto the transformed reciprocal space. One can easily image the richness of band diagrams in transformed periodic structures that would be controlled by the function f. Interesting extensions of classical mathematical theorems and their applications to physics of transformation crystallography will be addressed elsewhere.

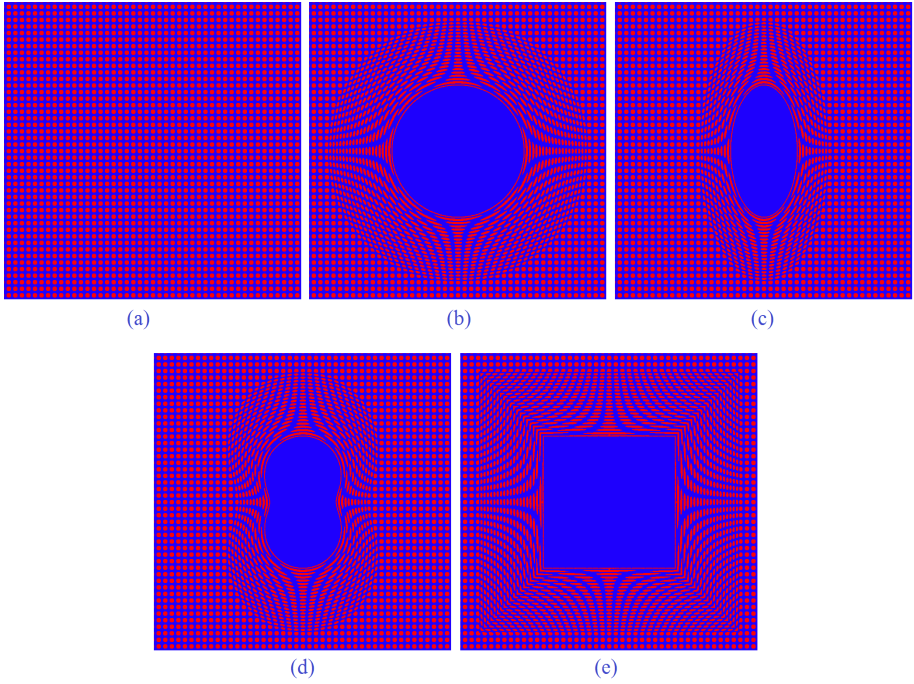

Figure 2: Transformed square lattices (four-fold): (a) square lattice; (b-e): transformed square lattice with a circular cloak (b), an elliptic cloak (c), a peanut cloak (d), a square cloak (e).

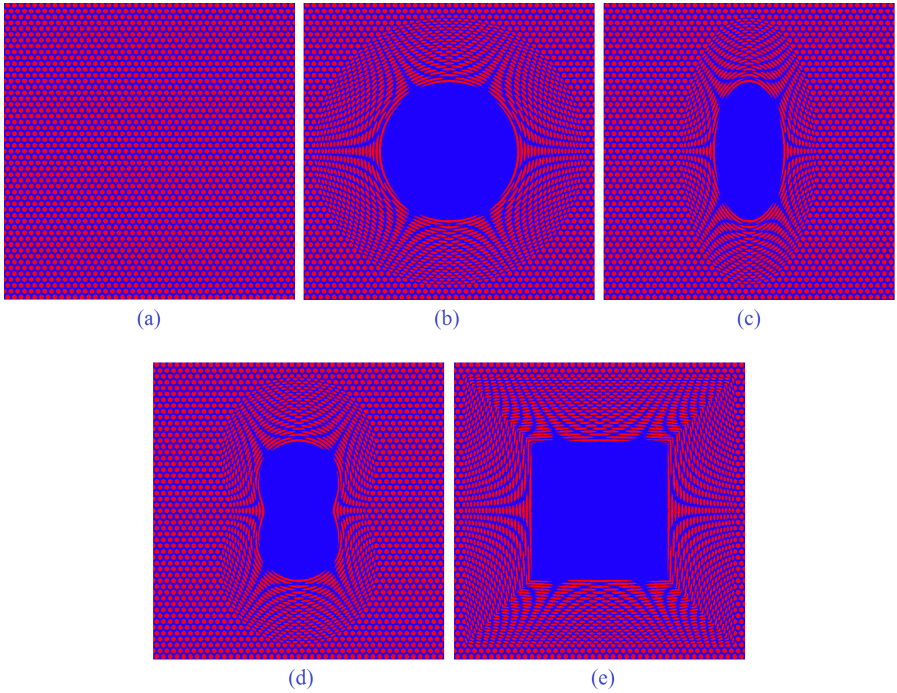

Figure 3: Transformed triangular lattices (threefold): (a) square lattice; (b-e): transformed triangular lattice with a circular cloak (b), an elliptic cloak (c), a peanut cloak (d), a square cloak (e).

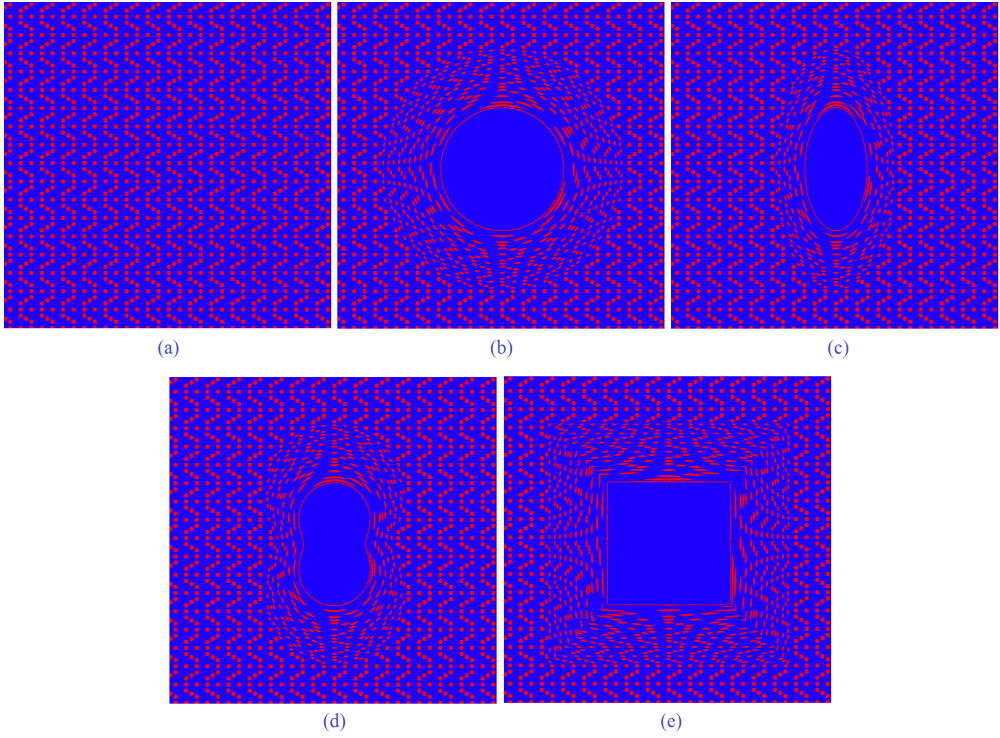

Figure 4: Transformed pentamode-like lattices (two-fold): (a) pentamode lattice; (b-e): Transformed pentamode lattice with a circular cloak (b), an elliptic cloak (c), a peanut cloak (d), a square cloak (e).

## 3. Numerical illustrations with FDTD

The computer software SimSonic, which we use in this section to simulate propagation of elastic waves in crystals and transformed crystals is a freely available software designed by Emmanuel Bossy [14], which is based on finite-difference time-domain (FDTD) computations of the elastodynamic equations. SimSonic serves as a tool for communities of researchers, teachers and students. The SimSonic package consists of several compiled programs and C source codes, freely available, under the GNU GPL license[14]. SimSonic solves the Navier equations in the following form:

$$\frac{\partial T_{ij}}{\partial t}(x,t) = \sum_{j=1}^{d} \sum_{i=1}^{d} C_{ijkl}(x) \frac{\partial v_k}{\partial t}(x,t) + \theta_{ij}(x,t)$$

$$\rho(x) \frac{\partial v_i}{\partial t}(x,t) = \sum_{j=1}^{d} \frac{\partial T_{ij}}{\partial t}(x,t) + f_i(x,t)$$

where $C_{ijkl}$ is the rank-four symmetric elasticity tensor with $2^4$ entries, $T_{ij}$ the rank-2 stress tensor and $v_k$ the particle velocity (vector) field. Besides from that, $\rho(x)$ is the density, $f_i(x,t)$ and $\theta_{ij}(x,t)$ the force (vector field) and strain rate (rank-2 tensor) sources, respectively, with $x=(x_1,x_2)$ is the space variable and $t>0$ is the time variable, and the dimension of the computation $d=2$.

In Figure 5, we show some snapshots of SimSonic computations that reveal the wave pattern of a seismic wave propagating within a graphene-like seismic metamaterial (a-c) with geometric parameters as in Figure 1(a) and the same seismic propagating in the graphene-like seismic metamaterial after a stretch of coordinates has been made to achieve an elliptical cloak as in Figure 1(c). It can be noted that this creates an exclusion zone in the center of the elliptical cloak, where the seismic wave magnitude is reduced by almost one half. For the bulk medium (soft soil) parameters, we consider a density of $10^3/m^3$ whereas the density of inclusions is $10^4/m^3$ (denser soil). The coefficients of the elasticity tensor used for the bulk medium (e.g. soft soil with an assumption of isotropic homogeneous medium) are

$C_{11} = C_{22} = 0.25$ GPa, $C_{12} = 0.25$ Gpa, $C66 = 0$,

The coefficients for the elasticity tensor used for the inclusions (e.g. concrete with an assumption of isotropic homogeneous medium) are

$C_{11} = C_{22} = 25$ GPa, $C_{12} = 25$ GPa, $C66 = 0$,

Computations are carried out on a square region 225 x 131 $m^2$ i.e. with 50 x 50 hexagons with circular inclusions located at each vertex with 1.5 m as center-to-center spacing and with a radius of 0.5 m. Each simulation runs for 1000 ms with a snapshot taken every 10 ms. The computational domain is a rectangle of side lengths 898.2 m by 524 m.

With all these assumptions, we satisfy the stability condition (CFL condition, from the initials of Courant, Friedrichs and Levy) for the numerical scheme:

$$\Delta t \leq \frac{1}{\sqrt{d}} \frac{\Delta x}{c_{max}}$$

where d=2 is the space dimension, $\Delta x = 0.2$ m and $\Delta t = 10$ μs are the space and the time discretizations, $c_{max}=(C_{11}/\rho)^{1/2}=(25 \ 10^9/10^4)^{1/2}=1580$ m/s is the related to the speed of pressure waves in concrete.

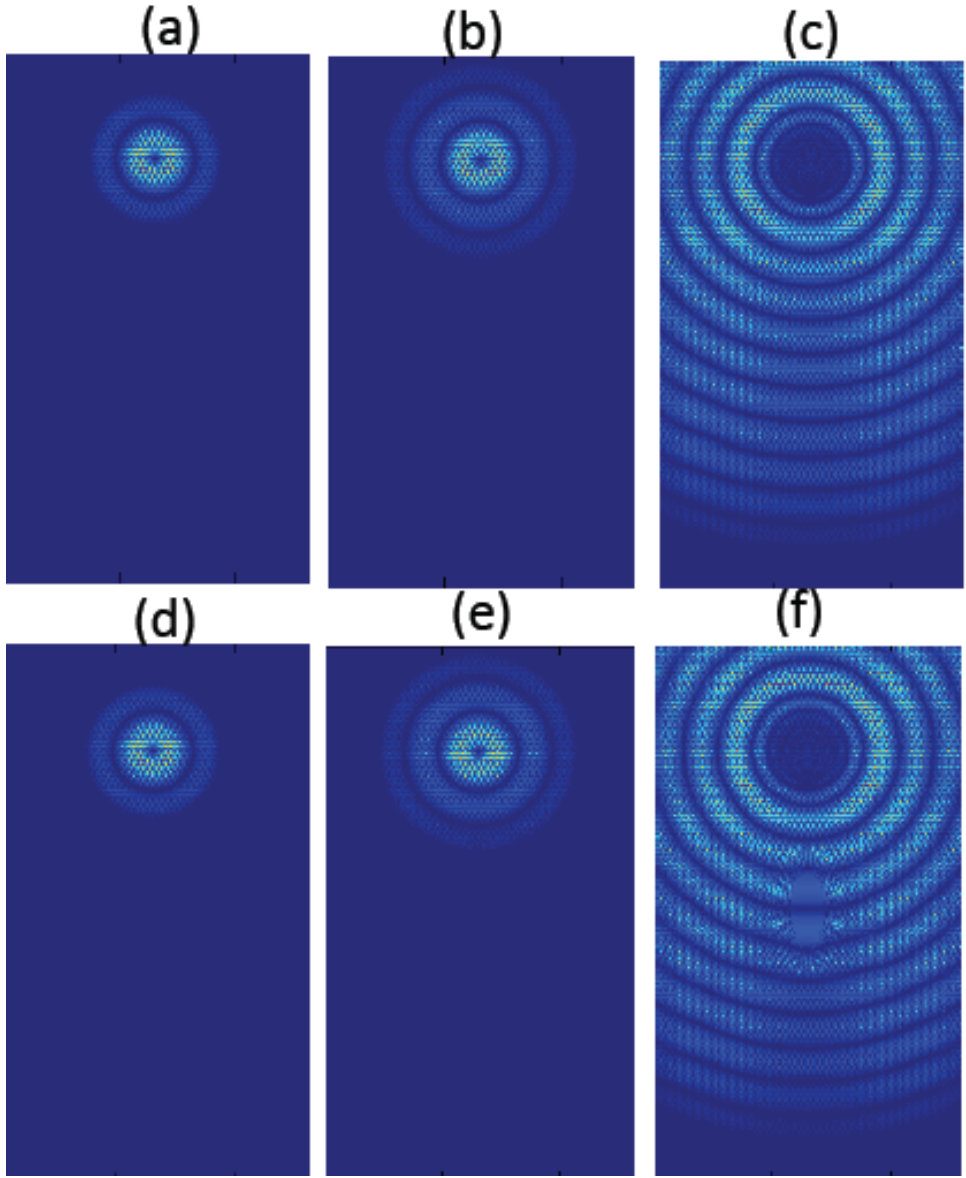

Figure 5: Numerical simulations (SimSonic software) for a point force oscillating out-of plane ($x_3$-direction orthogonal to ($x_1,x_2$) plane of computation) at frequency 10 Hz generating a seismic wave propagating in a graphene-like seismic metamaterial (soft soil with columns of denser soil) like in Figure 1(a) (upper panel) and the same with an elliptical cloak like in Figure 1(c). Snapshots at t = 100 ms (a,d), t = 150 ms (b,e) and t = 570 ms (c,f). Note the reduced wave amplitude in the center of the cloak in (f) compared to (c). The center to center spacing of columns is 1.5 m and their diameter is 0.5 m in (a,b,c), while in (d,e,f) the mapping has stretched these distances like in Figure 1(c). Linear color scale ranges from dark blue (vanishing) to yellow (maximum) displacement field.

In Figure 6, we show similar snapshots for square (a), circular (b) and peanut (c) seismic cloaks at 570 ms. In all three cases the displacement field has smaller amplitude within the exclusion area (seismic protection), about half of the seismic wave energy is smoothly detoured around the cloak. Importantly, there is virtually no reflection of the seismic wave by the cloak.

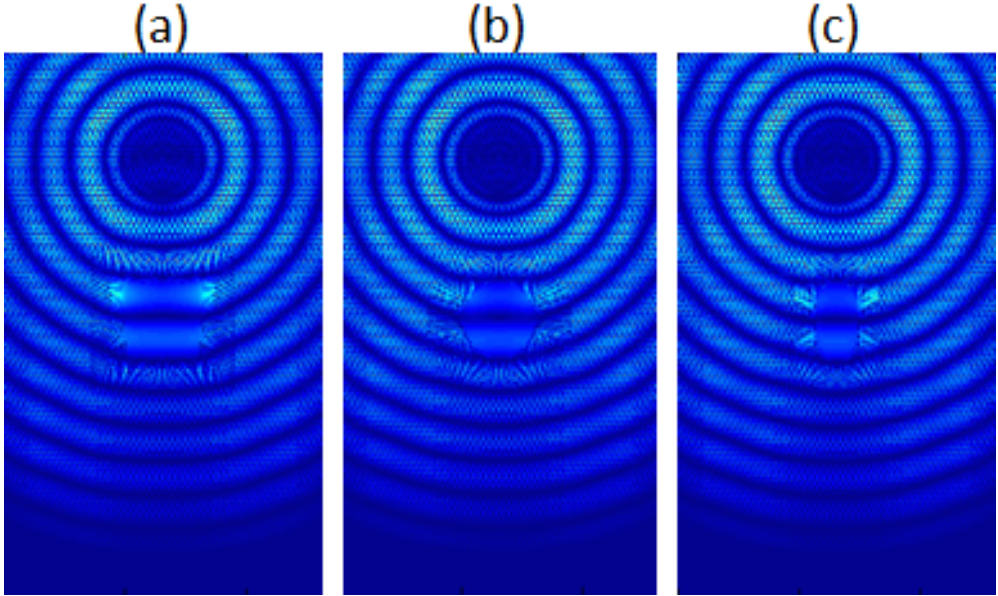

Figure 6: Numerical simulations (SimSonic software) for a point force oscillating at frequency 10 Hz generating a seismic wave propagating in a graphene-like seismic metamaterial (soft soil with columns of denser soil) like in Figure 1(e), a square cloak (a), Figure 1(b), a circular cloak (b), and Figure1(e), a peanut cloak (c). Snapshots are taken at t = 570 ms. Linear color scale ranges from dark blue (vanishing) to red (maximum) displacement field.

In Figure 7, we propose a concept of Penrose-like seismic cloak, which is based upon a geometric transform in a Penrose lattice. The latter has been designed using cut-and-projection method, as discussed in section 2. However, we numerically observed that the level of protection displayed by this cloak is less prominent than for the graphene-like cloak in Figures 5 and 6 for the same source oscillating at 10 Hz, so we report here the result of SimSonic computation when the source oscillates (out of plane i.e. along $x_3$) at frequency 20 Hz, in which case seismic protection is achieved in the $(x_1,x_2)$ plane (in-plane seismic signal). We nevertheless are convinced that transformed quasi-periodic lattices offer a very promising route towards seismic cloaks, as one can generate such lattices from cut-and-projection of periodic lattices in higher-dimensional spaces (5 and 12 for instance) and then further choose the transform allowing for designs of quasi-periodic lattices (with 5-fold and 12-fold symmetries for instance) with an exclusion area of any given shape. Such transformed quasi-periodic lattices are reminiscent of Vasarely's artwork[11], although the artist used geometric transforms in periodic rather than quasi-periodic lattices.

Let us now move to the morphing technique, which can be viewed as a geometric transform, although in the present case, it is based on control points rather than an explicit mathematical formula, so one might be tempted to call this an "inverse engineering geometric transform".

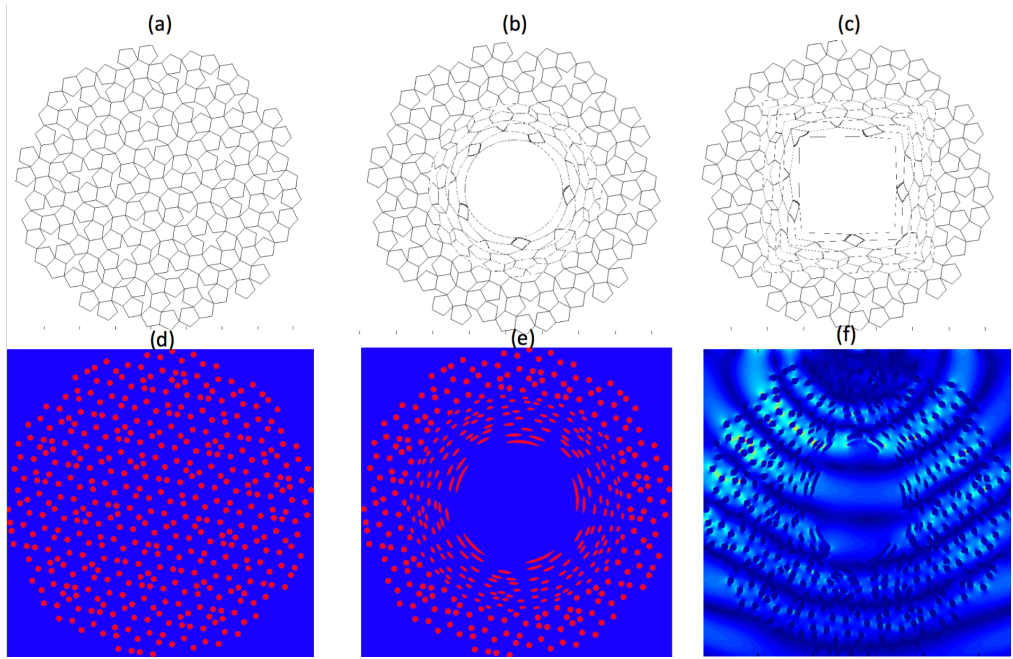

Figure 7: Transformed Penrose lattices (five-fold symmetry): (a) Penrose lattice from cut-and-projection method; (b-c) Transformed Penrose lattice with a circular cloak (b) and a square cloak; (d-e) Same with vertices only; (f) Numerical simulation (SimSonic software) for a point force located just above the cloak and oscillating at frequency 20 Hz along $x_3$ generating a seismic wave propagating for Penrose-like seismic metamaterial (soft soil with columns of denser soil). Linear color scale ranges from dark blue (vanishing) to red (maximum) displacement field.

## 4. Morphing

The early late 80's saw a rise in the popular media use of the effect called "morphing" which consisted of transforming an image into second image with a succession of intermediate images. This computer graphic technique, notably used in the heroic fantasy movie *Willow* and the musical video clip *Black or White* of Michael Jackson could be helpful in other contexts, such as in transformation physics and wave simulations. Indeed, the morphing approach is reminiscent of transformational optics, whereby a coordinate change stretches a Cartesian grid (or more generally a Delauney mesh) in a way similar to what researchers use to control light trajectories in transformed (metamaterial) media[16]. In following paragraphs, we would like to show that morphing can deepen our understanding of transformational crystallography by adopting a different viewpoint. Actually, of all the sciences, computer science seems to be the most frequently employed on a regular basis by researchers from miscellaneous fields, including photonics and biological sciences. It can be also helpful for people working on transformation physics. Indeed, by combining physical and computer sciences, we achieve in the sequel unique applications of morphing techniques.

*4.1 Principle of morphing*

Morphing literally speaking is an image transformation. If at the beginning, the morphing was just a simple interpolation of colors, with the source's image colors, which progressively fade away to allow for the appearance (also progressively) of the destination's image colors, we should stress that nowadays morphing is slightly more complicated. In fact, morphing is now based on a double interpolation, both on shapes and on colors, between two images. There are several methods to do this, and we have chosen to investigate the most wide-spread methods. To illustrate our discussion, we have used freely available software called, Sqirlz Morph [15].

*4.2 On the usefulness of control points*

Independent of the choice of morphing technique used, « control points » need to be selected manually and assigned by the user, in both source and destination images, to determinate the most important features of the images. The user has to make sure that each control point in one image corresponds to one control point in the second image, in order to establish a one-to-one mapping. Control points are essential for the shape's transformations. Indeed, since morphing preserves the proportions, if there are some parts of one image placed at the right position in the other image, there is be a higher probability that the whole transformation would be correct. In fact, a higher number of control points means a higher probability of success. However, control points should be neither aligned nor in too close proximity to each other. Their number is also important. Indeed, too few control points will produce a superposition of the source and destination images, instead of a real transformation of one into the other. On the other hand, too many control points might produce antagonistic transformations and ruin the final result. Therefore, choosing the location of the control points requires consideration of many parameters, which means that this step is not easily amenable to automation [4]. This is in pivotal drawback in being able to achieve faster results: Any human intervention means a more subjective result.

*4.3 Application to design of seismic metamaterials*

Let us consider panels (b)-- a circular cloak-- and (b)—an elliptical cloak-- in Figure 2 and apply the morphing algorithm. We emphasize the importance of control points in Figure 2(b,c), which leads to Figure 8(a). The reader can easily create other designs with Sqirlz Morph. It should be noted that in two extreme cases whereby the number of control points is either far too small or far too large the result of morphing can be quite surprising. By doing so one can generate beautiful patterns that can serve the purpose of bio-inspired seismic metamaterial or as aforementioned artwork reminiscent of that of Vasarely [11]. In the present case, we have designed a seismic cloak, using Pendry's transform [6,7,9,10], which serves the purpose. One can see that among the three snapshots taken from Simsonic simulations of an elastic wave propagating in soft soil (blue) with denser inclusions (red), the last one (d) at 520 ms shows a reduction of the elastic field magnitude in the exclusion zone of the cloak. We have checked that this remains the case at larger time steps. However, at short times, it is clear from (b) and (c) that the cloak is not yet efficient for protection. Nevertheless, one notes that the elastic field displacement is nearly constant in the exclusion zone and the soil therein moves almost like a rigid body, which suggests some kind of trapped fundamental mode of a stress-free cavity. Importantly, invisibility is already achieved at short times: a velocimetre placed behind the cloak would not detect any significant change of the seismic signal compared with the case where there is no cloak: the wavefront is almost circular and its amplitude is close to that of the wave propagating in the medium shown in Figure 2(a) i.e. without cloak.

It is illuminating to compute the difference between 75% morphing (Figure 8(a)) and direct geometric transform between Figures 2(b) and 2(c) when we consider an elliptical cloak of eccentricity ¾ to that in Figure (c). Figure 8(e) shows that difference in $L^2$ norm computed with the formula $F = \sqrt{\frac{1}{N} \sum_{i=1}^{N} \left(\frac{K_i}{255}\right)^2}$ (where the sum is taken over all the pixels $(K_i)_{i=1,...,N}$ in the image and each pixel $K_i$ has a value between 0 and 255) is less than 1% (for images converted in grayscale). Importantly, this formula only works for grayscale images (otherwise, one would have to consider a vector valued function $K_i$ with 3 components for Red, Green and Blue, respectively). Interestingly, the structural similarity (SSIM) index[36], which is a method for predicting the perceived quality of digital images gives a difference of almost 45%. Actually, one can see that Figure 8(e) is almost black (vanishing $L^2$ norm), whereas Figure 8(f) clearly has striking instantly recognizable differences. SSIM is based on a complex mathematical formula that can be found in[36], but roughly speaking it is a perception-based model that considers image degradation as perceived change in structural information, while also incorporating important perceptual phenomena, including both luminance masking and contrast masking terms.

We do not find such antagonistic results between $L^2$ norm and SSIM in our previous investigations of morphing applied to transformation optics[16], which were based on coordinate stretches in invisibility cloaks surrounded by homogeneous medium. In the present case, the homogeneous bulk is replaced by a periodic cladding, so the coordinate stretch is much more challenging to mimic with morphing (as one has to place many control points in the periodic cladding). We believe we are in presence here of configurations where these two estimates for image differences ($L^2$ norm and SSIM [36]) cannot be reconciled (a whole class of transformational crystallography based images would indeed face the same fate). However, regarding numerical simulations, it is usual to use $L^2$ norm for error estimates, and we have checked that FDTD simulations of elastic waves propagating in morphing based medium in Figure 8(a), see snapshots in Figures 8(b) and 8(c), and corresponding snapshots for elastic waves propagating the transformed medium from 75 percent stretching in Figures 2(b) and 2(c), are almost identical (less than one percent of difference in $L^2$ norm).

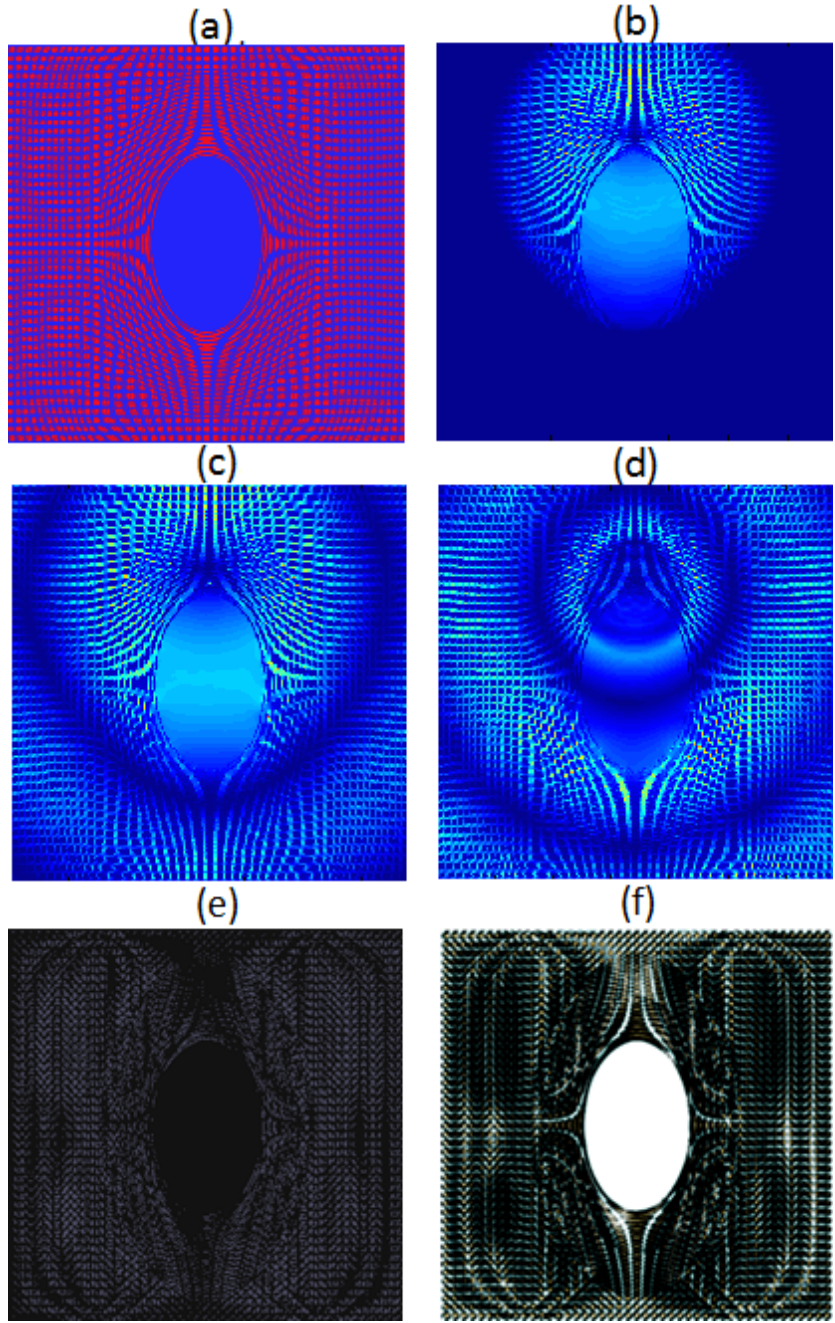

Figure 8: Bio-inspired cloak generated by 75% morphing from Figure 2(b) to Figure 2(c); (a) is the resulting stretched image; (b-d) Snapshots at 100 ms (b), 200 ms (c) and 520 ms (d) from Simsonic computations for a source at frequency 20 Hz. Blue pixels for soft soil and red pixels for denser soil. Note that protection becomes more effective at 520 ms. (e) Difference in gray scale between Fig. 8(a) and elliptical cloak with 75% of eccentricity of that in Fig. 2(c) corresponding to 0.0849 in L2 norm; (f) Same as (e) for SSIM index[36] that gives 0.4461 of image difference. We point out that for error estimates in numerical results, the L2 norm is the natural tool of choice.

## 5. Graphene flakes as bio-metamaterials for control of mass diffusion

In this section, we aim to begin to bridge concepts of seismic wave physics and diffusion phenomena, in the present case mass diffusion (note that mass diffusion is governed by a Fick's equation whereas heat diffusion is governed by Fourier's equation), to enlarge our horizon on the design of cloaks. In Figure 9, one can see two numerical simulations in the upper panel, placed next to each other in order to emphasize the similarities (and differences) between the wave and diffusion patterns. In both cases, a layered cloak simultaneously leads to invisibility and protection. From the seismic wave pattern in Figure 9(upper left), it appears clear that the center of the cloak is the ideal location to build a monument: the wave magnitude vanishes there. Similarly, from the chemical concentration distribution in Figure 9(upper right), one could envision some delayed drug diffusion if the drug is placed in the center of the biocloak: the inner boundary of the cloak acts like a barrier for chemical species, it is difficult to get inside the invisibility region and by reciprocally the drug can be concentrated in this region.

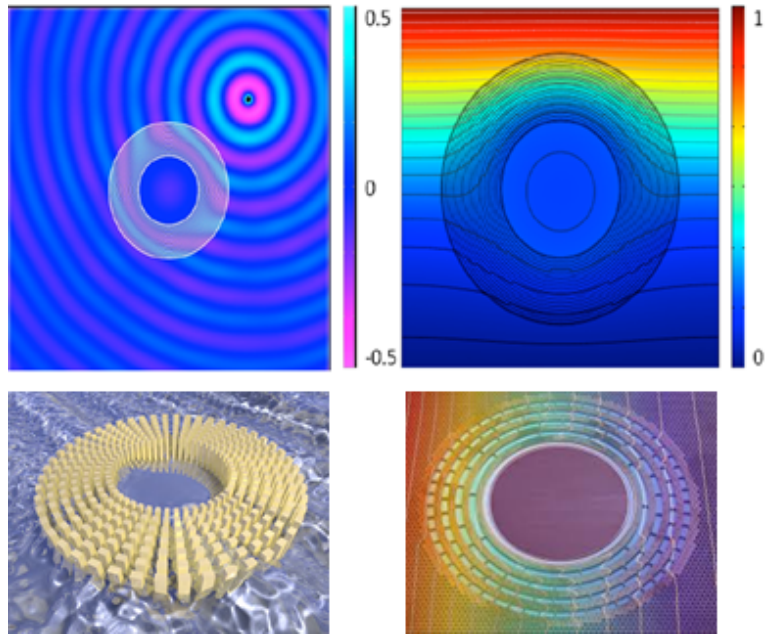

Figure 9: Mechanical versus thermal cloaking for invisibility and protection; (Upper panel) Numerical simulations for the propagation of a Rayleigh-like wave in a plate of thickness 40 m generated by a time-harmonic point source of frequency 10 Hz located in the close vicinity of a layered cloak (based on a scaled up version of cloak by Mohamed Farhat et al. [27]) of diameter 500 m (upper left) and diffusion of heat (from top to bottom) in a layered cloak (upper right) of diameter 10 cm; (Lower panel) Artistic view by Tolga Ergin (Martin Wegener's team at KIT) of ocean wave cloak (based on a scaled up version of cloak by Mohamed Farhat et al. [26]) (lower left) that works in the same manner has the earthquake cloak (rigid pillars have the same geometry whether they are plunged in soil or water) and result of Robert Schittny (Martin Wegener's team at KIT) experiment on control of heat diffusion in false colors (lower right); Note that the wave amplitude (left) and heat magnitude (right) vanish in the center of the wave and diffusion cloaks.

These two cloaks, which are of very different scales (tenths of meters for the seismic cloak and tenths of nanometers of the drug diffusion cloak), have been somewhat already experimentally validated, see lower panel for an illustration by Tolga Ergin (group of Martin Wegener at Karlsruhe Institute of Technology, that also produced a magnificent computer based photorealistic view of negative index materials [21]) of the protection offered by an ocean cloak, which is a scaled up version of the water wave cloak designed and tested experimentally in [26].

The experimental measurements of Robert Schittny (Wegener's group at KIT) for a thermal cloak studied in the time domain in 75, see lower right panel in figure 9, can be used to foresee what can be achieved for control of drug diffusion with concepts developed in [34]. In the same way, if one finds it legitimate to draw analogies between water waves (ocean cloak), surface Rayleigh waves (seismic cloak) and heat and mass diffusion (biophysics cloak), the latter being solutions of the Fourier-Fick equation:

$$\sum_{i=1}^{d} \frac{\partial}{\partial xi} \sum_{j=1}^{d} \kappa jk(x,t) \sum_{k=1}^{d} \frac{\partial}{\partial xk} T(x,t) = \frac{\partial}{\partial t}(\rho(x,t)c(x,t)T(x,t))$$

Where κij is the conductivity tensor, and ρ and c represent respectively the density and specific heat in the case of heat, a product which is equal to 1 for the case of mass diffusion, then another interesting parallel can be drawn between the biophysics at the nanoscale and the geophysics at the meter scale. At a nanometer scale, the diffusion of different type of macromolecules in matrix-based context is highly studied and is the basis for important biological measurements techniques such as fluorescence recovery after photobleaching (FRAP) and fluorescence correlation spectroscopy (FCS) [37,38]. It has recently been shown that the structures of materials such as graphene and derivatives of graphene, graphene oxide can be exploited as molecular filters allowing the permeation of ions of certain hydrated radii to pass through [39,40]. Matrix structures have already been describing which combine graphene and clay based structures, which both can be modelled to use condensed matter theory and effective medium approaches to describe the movement of aqueous media [34,41,42,43]. Graphene-matrices using therapeutic molecules for human therapy have also been developed. Though graphene materials were initially used simply for chemical conjugation purposes, it has now been shown that graphene is able to produce retardation effects which are useful in the development of controlled or slow-release drug formulas [34,44]. This is an important health application as controlled release drugs have the advantages of increasing the plasma half-life of therapeutic drugs which are associated with economic benefits such as fewer injections for patients [45] and corresponding reduced healthcare visits.

Upon inspection of Figure 10, it can be noted that the house of card type configuration of graphene oxide flakes from transmission electron microscopy (a), which are in the scale of a few micrometers is similar to layers of clay in soils, which are in the scale of a few metres. Following image treatment with Matlab (b), one can export the house of cards geometry in Simsonic and simulate the propagation of a seismic signal upon change of scale by a factor $10^6$. The house of card configuration in Figure 10(a) can be exploited to build an effective porous diffusion model for drug diffusion through a graphene oxide membrane as achieved in [3] using mathematical analogies with diffusion models in porous soils. However, one can see in the present case that upon scaling by a factor one million, and change of the elastic properties for that of clay in soft soil, a seismic lens can be achieved, see Figure 10(d). One can attribute this elastic field localization to a thin bridge of soft soil (white) between two denser regions of soil (black) at the location of the image, by comparisons of panels (b) and (d). Indeed, it is well known that fields oscillate faster in thin domains, see [26] and references therein. We believe that interplay between models and experiments in geophysics and nanoscale biophysics could lead to major discoveries in the near future.

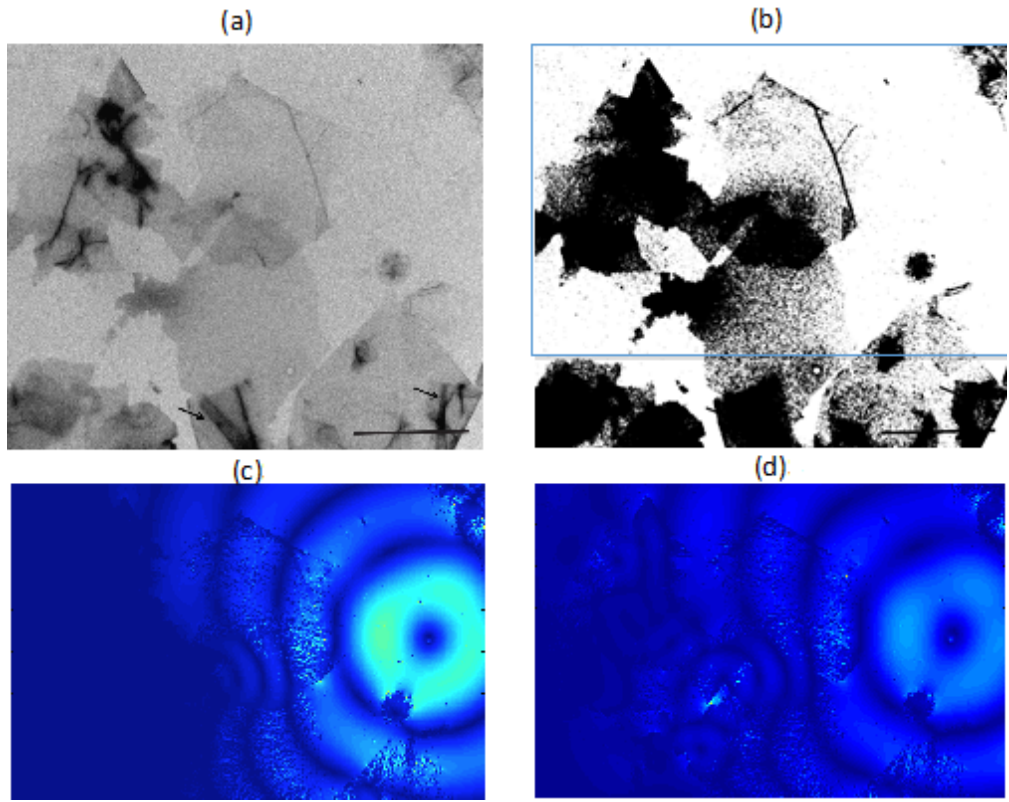

Figure 10: Bioinspired seismic lens from TEM images: (a) Transmission electron microscopy (TEM) images of a modified form of graphene, graphene oxide (GO). Images are acquired using a Morgagni FEI 80KV Camera digital View III Olympus camera. Multiple sheets show a sharp increase in the contrast when many GO flakes are stacked together, thereby substantially increasing the gray scale of the images. Scale bar (2 μm). Black arrows represent regions displaying the property of high flexibility of the GO sheets, which are capable of folding over though remaining intact. (b) Same image after treatment, with white pixels for soft soil and black pixels for denser soil (we select a region of the image for numerical simulations, marked in blue with 1139 pixels x 750 pixels and each pixel is 0.2 m x 0.2 m) ; (c-d) Snapshots at 125 ms (c) and 250 ms (d) from Simsonic computations for a source at frequency 20 Hz. One notes the focusing of seismic wave in (d).

## 6. Field test experiments in bio-inspired seismic metamaterials

Let us now investigate the similarities between models of nano-scale photonic and phononic crystals, and metre-scale seismic metamaterials, as introduced in recent papers[4]. We recall in Figure 11 the scheme of the first large scale experiment on seismic metamaterials (left panel), which was conducted by the dynamic soil laboratory team of Stéphane Brûlé at the Ménard company in August 2012. The experimental data (right panel) shows that when a seismic source oscillates at a frequency inside the stop band of the periodic structure, here 50 Hz, Rayleigh waves get partly reflected, just like a photonic crystal does for light.

   A keen observer of civil engineering works can observe that certain artificial and buried structures in the soil seem to be a translation of crystallographic lattices (Figure 12). The common structure is made of vertical and cylindrical inclusions (concrete, steel, etc.) implemented in the soil, reproducing a square meshing in the plan (0, x, y). To achieve an improvement in the density of soil strengthening, the equivalent of a face-centered cubic system can be observed too in plan (0, x, y) with vertical concrete rods.

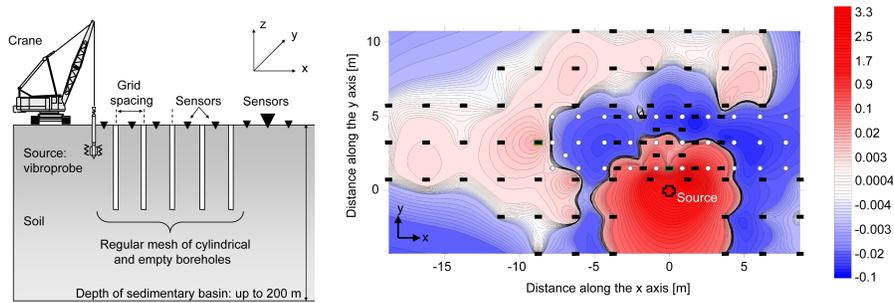

Figure 11: Scheme of the 2012 field test experiment conducted by the Ménard company in Grenoble (left panel): An oscillating probe generates acoustic waves at 50 Hz in front of a mesh of cylindrical boreholes. An array of sensors monitors the intensity of the waves at various positions. Experimental results (right panel): The map (black rectangles: sensors; white circles: boreholes; red cross: source) plots the difference of energy after and before drilling the boreholes. The dark blue region behind the holes indicates that the presence of the borehole mesh results in a decrease of elastic energy transmitted to that area. The red region indicates the area close to the source in which the wave intensity increases because of reflection. Credit: S. Brûlé et al. [4].

However the primarily objective of these worksites is to make the soil more resistant to shear strains induced by a simplified representation of the seismic disturbance, i.e. considering that the more devastating component of the seismic signal correspond to the horizontal component of the body shear-waves. The fascinating concept carried out by several authors and that has boomed in the past few years is the introduction of the concept of photonic and phononic crystals in geophysics. As aforementioned, in August 2012, a first full-scale field experiment was realized with a non-sub wavelength 2D square grid of vertical empty cylindrical holes disturbed by a 50 Hz source, showing a characteristic Bragg's effect[4] reminiscent of stop band properties of photonic crystals [46,47] put forward by Yablonovitch and John in 1987. Such periodic dielectric structures, that were also studied by Stokes and Rayleigh towards the end of the nineteenth century [48], and by Bykov and Ohtaka in the seventies [49,50], allow for spontaneous emission of light and almost total reflection of light. The idea proposed, and experimentally demonstrated in [4], is to simply scale up photonic crystals (which have typically an array pitch on the order of a few hundredth of nanometres) by a factor of $10^4$ to $10^5$, so as to reflect surface seismic wave a few metres in wavelength in structured sedimentary soils.

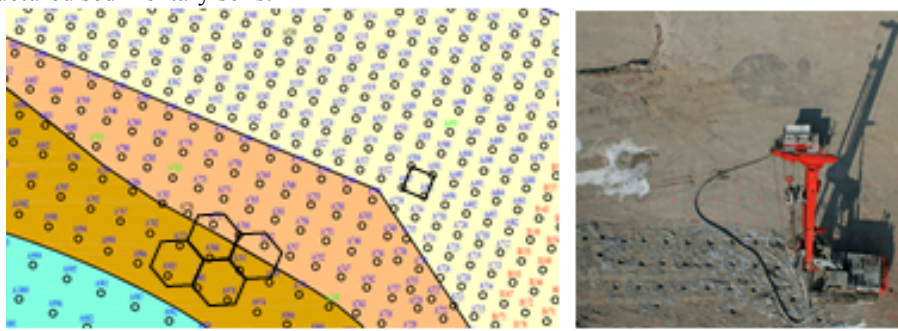

Figure 12: Illustration of high-density ground reinforcement with vertical, cylindrical lime inclusions arranged in a square grid (right panel). The horizontal spacing is comprised between 1.3 and 2.0 m and the diameter of the inclusions is around 0.3 m (courtesy of Ménard). Left, the nodes of the grid could be design in agreement with the graphene-sheet structure.

Though this idea initially appears as far-fetched, leading to initial skepticism from certain research groups, it is now appears to be widely accepted, and further, experiments have been by the geophysics team of Philippe Roux on stop band properties of forest of trees[51] in the tracks of large-scale seismic metamaterials and tested by Menard engineering group[4]. Another interesting way to protect an area is actually to convert surface Rayleigh waves into shear bulk waves, which is precisely what the metawedge does [52].

In the present work, we emphasize that in the near future, it would be interesting to identify the possibilities to value the properties of the graphene-sheet structure in civil engineering. Indeed, graphene has a unique band structure with so-called Dirac cones, which are frequencies where the periodic structure behaves like an effective medium with less than ordinary properties, such as a near-zero refractive index allowing for Dirac cone cloaking[53-58]. Thus far, only transmission properties have been studied near Dirac cones, but there may be well also protection features.

Apart from these features, the soil-liquefaction remediation in case of earthquake could be achieved by incorporating a mesh of vertical concrete walls (shear-walls), or jointed piles or inclusions in the soil as depicted in Figure 13 (left). One can imagine to reproduce thus the graphene structure. It can be noted that in 2002, Takemiya and Shimabuku suggested the realization of a "semi-honeycomb pattern" around viaduct piles (Figure 13, right) with soil-cement mixed columns. Their objective was to improve the global stiffness without immensely modifying the mass of the system. These buried structure are expected to work not only for reducing the seismic input thanks to the cell stiffness but also for absorbing seismic energy by self-damaging in case of severe earthquake.

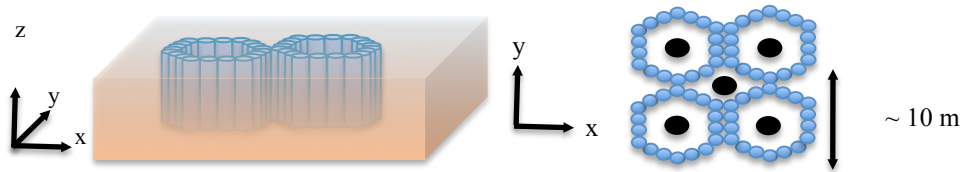

Figure 13: Principle of a honeycomb (left) and "quasi-honeycomb" (right) meshes made of jointed concrete inclusions (blue) and viaduct piles (black) in the soil (brown). Piles and soil-cement mixed columns[35] would act as protection for viaduct piles.

## 7. Concluding remarks

In this review article, we have explained how geometric transforms and morphing work and their concepts can be applied to different contexts, such as transformation crystallography, and quasi-crystallography (for which we have recalled the method of cut-and-projection to generated ad libitum quasi-periodic structures). We have hope to have convinced the reader that crystallography is the essence of bio-inspired seismic metamaterials. Notwithstanding the 1888 theoretical contribution of Lord Rayleigh to crystallography [48], it is widely accepted that this science was boosted over a century ago, by the British physicist, chemist and mathematician Sir William Henry Bragg and his (Australian born) son Sir William Lawrence Bragg. These pioneers in surface science shared the 1915 Nobel Prize in Physics for their discovery of Bragg's law of X-ray diffraction[79], $2d \sin \theta = n\lambda$ for constructive interference within an atomic lattice (where d is the atomic spacing in the crystal lattice, n is a positive integer, $\theta$ the scattering angle and $\lambda$ the wavelength of incident wave, which is typically a few angstroms like atomic bonds, so three orders of magnitude shorter than that of light).

Interestingly, Sir William Lawrence Bragg is the youngest physicist to have received the Nobel Prize (he was 25 years old in 1913) and he was head of the Cavendish laboratory when the discovery of structure of DNA (deoxyribonucleic acid) was reported by the American biologist James Dewey Watson and his British molecular biologist colleague Francis Crick in 1953. DNA discovery was made possible thanks to scientific discussions with Rosalind Franklin, an English chemist and X-ray crystallographer, who suggested the double helix nature of DNA while working at King's College London.

One can therefore see that crystallography and biology have had a long lasting fruitful joint history, which is certainly due to the similar scales involved (a few nanometers down to angstroms). Graphene, photonic crystals and indeed metamaterials have renewed the interest in crystallography, the former since it (re)opens a door to material science at sub-nanometer scale, the latter since Bragg's scattering is intimately linked to stop bands. However, when viewed from the sky, soils structured at the meter and decameter scale (either through manmade civil engineered techniques or simply arising from in nature, like with forests of trees) display a geometry akin crystals. We therefore proposed in this article to use the terminology *bio-inspired seismic metamaterials* to point out the rich history behind the young science of large scale mechanical metamaterials.

We also performed some numerical simulations in elastodynamic wave physics with some mention of applications in mass diffusion, a concept put forward by two of us in [34], which is supported by experiments for an extreme control of heat diffusion process with a thermodynamic cloak, which looks strikingly similar to the mechanical cloak for Lamb waves in plates [59]. A concise graphical summary of the power of morphing is depicted with the comparison of the images in Figure 8: The left and right images are virtually indistinguishable by the naked eye from a distance i.e. one might think that the corresponding effective media should share many features. However, this is the marked difference between a seismic metamaterial and a bio-inspired seismic metamaterial. The latter wishes to mimic nature's biological structures in order to reproduce some of the (many) interesting features that nature offers. For instance, the famous Morpho butterflies whose wings with an inner grating give their unique structural colors [60]. In the course of this article we have unveiled how morphing can be used to unveil new functionalities by mixing two known transformation-based metamaterials to get a new one which is reminiscent of a complex medium in living cells, which we coin as a bioinspired seismic metamaterial, inspired by the terminology bio-mimetism. Morphing may thus prove to be an invaluable tool for the exploration of transformation-based metamaterials from the nanoscale to the metre scale worlds. For instance, as first envisioned in a popular science oriented review paper[59], Martin Wegener proposed to cloak a city, which can certainly be achieved by using the exact same geometric transforms as in Figures 2-5, see Figure 14.

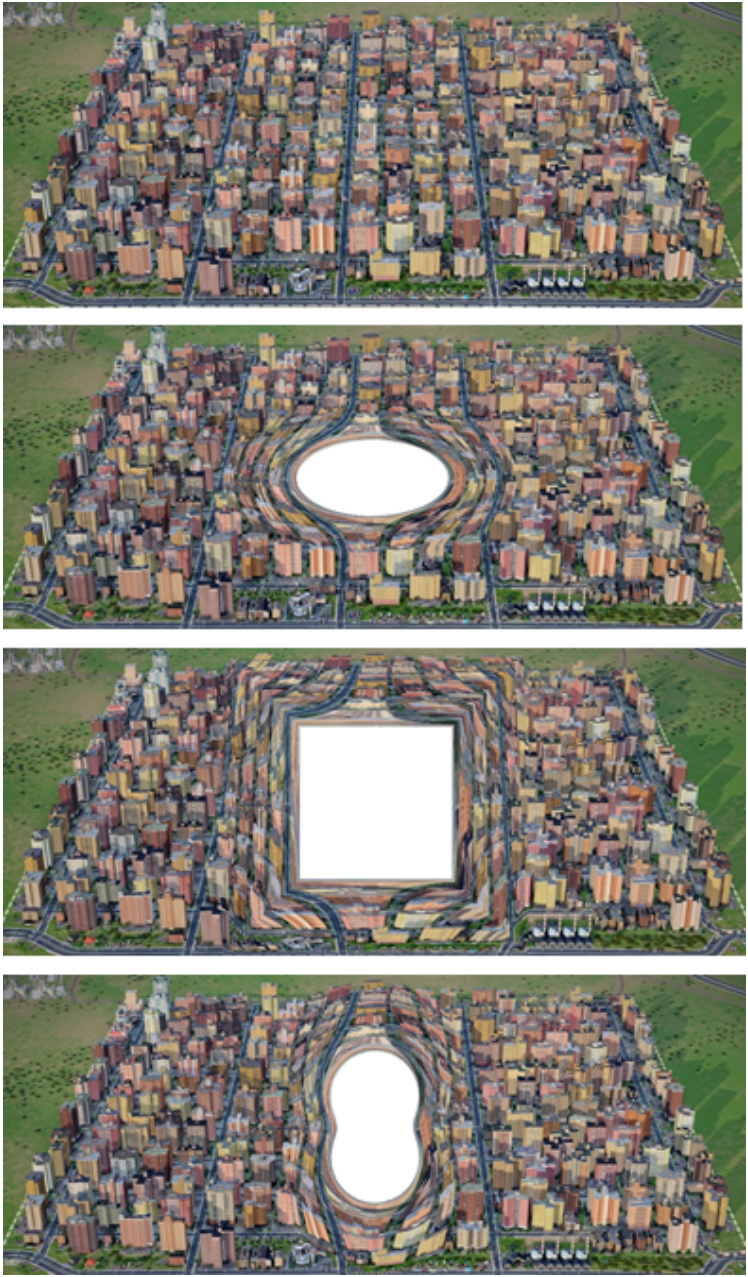

Figure 14: Geometric transforms applied to the schematic representation of a buildup city (a), could be used as a design tool for cities, with an exclusion zone of elliptical (b), square (c) and peanut (d) shapes. This exclusion zone could serve as a safety area for people living in the city in case of earthquakes, and could potentially be landscaped as a park with a lake. Inspired by Martin Wegener's group visionary infography[48].

The potential research area in seismic metamaterials is vast[4,32,61-73,] looking around for inspiration one notes notably beautiful quasi-periodic patterns in medieval architecture[64] and we have only touched upon the richness of cloaking phenomenon in the time domain[65]. As stated in[66]: *In the natural world, rays of light are sometimes bent when they pass through air layers of different temperatures, for example one of the most unusual mirages (called the* Fata Morgana *after the fairy Morgana from the Arthurian Legend) bends light rays in such an extreme way that on hot days boats appear to levitate above the sea [...] The beauty of Pendry's idea is that it allows matter to be engineered in such a way that light follows curved trajectories in metamaterials which are associated with geodesics in a transformed space. In the realm of transformational optics, one can envision many kinds of mirages, simply by distorting the space metric in what amounts to fabricating heterogeneous anisotropic media (also called metamaterials since they have been conceptualized and engineered by mankind).* We believe the exciting fields of transformation optics, acoustics, elastodynamics and crystallography are still in their infancy and will lead to unforeseen paradigms. Amongst them, urban and (possibly bio-inspired) seismic metamaterials[67-71] may prove an invaluable tool to design earthquake- resistant civil engineering cloaks in soft soils structured on the metre-scale by columns of concrete from the observation of nano-scale manmade or living organisms. Last, but not least, light, sound and heat interplay might form the essence of soon to come multiphysics small and large scale metamaterials based on geometric transforms that have proved their utilities for understanding the structure of the very small and very large worlds [80-82]. Transformation based metamaterials [83,84] lie somewhere in between these two frontiers (namely the scales of atoms and universe).

**Acknowledgements:**

R. Aznavourian, S. Guenneau and T. Puvirajesinghe acknowledge A*MIDEX project (no ANR-11-IDEX-0001-02) funded by the Investissements d'Avenir French Government program, managed by the French National Research Agency (ANR) with Aix Marseille Université (TMP, A_M-AAP-ID-14-15-140314-09.45-GUENNEAU-PUVIRAJESINGHE-HLS_SAT). S. Guenneau also acknowledges European funding through ERC Starting Grant ANAMORPHISM.

**References**

1. D. Shechtman, I. Blech, D. Gratias and J. Cahn, "Metallic Phase with Long-Range Orientational Order and No Translational Symmetry," Physical Review Letters **53** (20): 1951-1953 (1984).
2. K.S. Novoselov, A.K. Geim, S.V. Morozov, D. Jiang, Y. Zhang, S.V. Dubonos, I.V. Grigorieva and A.A. Firsov, "Electric Field Effect in Atomically Thin Carbon Films," Science **306** (5696): 666-669 (2004).
3. T.M. Puvirajesinghe, Z-L. Zhi, R.V. Craster and S. Guenneau, "Modulation of diffusion rate of therapeutic peptide drugs using graphene oxide membranes," arXiv preprint arXiv:1512.08506
4. S. Brûlé, E.H. Javelaud, S. Enoch and S. Guenneau, "Experiments on seismic metamaterials: Molding surface waves," Physical Review Letters **112** (13), 133901 (2014).
5. D. Terzopoulos, J. Platt, A. Barr and K. Fleischer, "Elastically deformable models," Computer Graphics **21** (4), 205-214 (1987).
6. J.B. Pendry, D. Schurig and D.R. Smith, "Controlling electromagnetic fields," Science **312**, 1780-1782 (2006).
7. F. Zolla, S. Guenneau, A. Nicolet and J.B. Pendry, "Electromagnetic analysis of cylindrical invisibility cloaks and mirage effect," Optics Letters **32**, 1069 (2007).
8. S-Y Lee, K-Y Chwa, J. Hahn and S.Y. Shin, "Image Morphing Using Deformation Techniques," The Journal of Visualization and Computer Animation 7 (1), 3-23 (1996).
9. A. Nicolet, F. Zolla and S. Guenneau, "Finite element analysis of cylindrical invisibility cloaks of elliptical cross section," IEEE Transactions on Magnetics **44** (6), 1150-1153 (2008).
10. A. Nicolet, F. Zolla, and S. Guenneau, "Electromagnetic analysis of cylindrical cloaks of an arbitrary cross section," Optics Letters **33**, 1584-1586 (2008).
11. http://www.vasarely.com/

12. V. G. Veselago, "The electrodynamics of substances with simultaneously negative values of ε and μ," Soviet Physics Uspekhi **10** (4), 509-514 (1968).
13. F. Guenneau, S. Chakrabarti, S. Guenneau and S. Anantha Ramakrishna, "Origami with negative refractive index to generate super-lenses," Journal Physics: Condensed Matter **26**, 405303 (2014).
14. E. Bossy, M. Talmant and P. Laugier, "Three-dimensional simulations of ultrasonic axial transmission velocity measurement on cortical bone models," Journal of the Acoustical Society of America **115**, 2314-2324 (2004).
15. http://www.xiberpix.net/SqirlzMorph.html
16. R. Aznavourian and S. Guenneau, "Morphing for faster computations in transformation optics," Optics Express **22**, 28301-28315," (2014).
17. T. Yang, H. Chen, X. Luo, and H. Ma, "Superscatter: Enhancement of scattering with complementary media," Optics Express **16**, 618545 (2008).
18. T. Bückman, M. Kadic, R. Schittny and M. Wegener, "Mechanical cloak design by direct lattice transformation," Proceedings of the National Academy of Sciences USA **112**, 4930-4934 (2015).
19. M. Rahm, D. Schurig, D.A. Roberts, S.A. Cummer, D.R. Smith and J.B. Pendry, "Design of electromagnetic cloaks and concentrators using form-invariant coordinate transformations of Maxwell's equations," Photonics and Nanostructures: Fundamentals and Applications **6** (1), 87-95 (2008).
20. Y. Luo, J. Zhang, J. Wu, Bae-Ian and H. Chen, "Interaction of an electromagnetic wave with a cone-shaped invisibility cloak and polarization rotator," Physical Review B **78** (12), 125108 (2008).
21. G. Dolling, M. Wegener, S. Linden, and C. Hormann, "Photorealistic images of objects in effective negative index materials," Opt. Express **14** (5), 1842-1849 (2006).
22. C.P. Berraquero, A. Maurel, P. Petitjeans, and V. Pagneux, "Experimental realization of a water-wave metamaterial shifter," Physical Review E **88**, 051002 (2013).
23. M. Brun, S. Guenneau and A.B. Movchan, "Achieving control of in-plane elastic waves," Applied Physics Letters **94**, 061903 (2009).
24. A. Diatta and S. Guenneau, "Controlling solid elastic waves with spherical cloaks," Applied Physics Letters **105**, 021901 (2014).
25. G. Dupont, O. Kimmoun, B. Molin, S. Guenneau and S. Enoch, "Numerical and experimental study of an invisibility carpet in a water channel," Physical Review E **91**, 023010 (2015).
26. M. Farhat, S. Enoch, S. Guenneau and A.B. Movchan, "Broadband cylindrical acoustic cloak for linear surface waves in a fluid," Physical Review Letters **101**, 1345011 (2008).
27. M. Farhat, S. Guenneau and S. Enoch, "Ultrabroadband elastic cloaking in thin plates," Physical Review Letters **103**, 024301 (2009).
28. G.W. Milton, M. Briane and J. Willis, "On cloaking for elasticity and physical equations with a transformation invariant form," New Journal of Physics **8**, 248 (2006).
29. J. Renger, M. Kadic, G. Dupont, S. Acimovic, S. Guenneau, R. Quidant and S. Enoch, "Hidden progress: broadband plasmonic invisibility," Optics Express **18**, 15757-15768 (2010)
30. N. Stenger, M. Wilhelm and M. Wegener, "Experiments on elastic cloaking in plates," Physical Review Letters **108**, 014301 (2012).
31. J. Xu, X. Jiang, N. Fang, E. Georget, R. Abdeddaim, J.M. Geffrin, M. Farhat, P. Sabouroux, S. Enoch and S. Guenneau, "Molding acoustic, electromagnetic and water waves with a single cloak," Scientific Reports **5** (2015).
32. P. Sheng, "Viewpoint: A Step Towards a Seismic Cloak," Physics **7**, 34 (2014).
33. D. Petiteau, S. Guenneau, M. Bellieud, M. Zerrad and C. Amra, "Spectral effectiveness of engineered thermal cloaks in the frequency regime," Scientific Reports **4**, 7386, (2014).
34. S. Guenneau and T.M. Puvirajesinghe, "Fick's second law transformed: one path to cloaking in mass diffusion," Journal Royal Society Interface **10**, 20130106, (2013).
35. H. Takemiya and J. Shimabuku, "Application of soil-cement columns for better seismic design of bridge piles and mitigation of nearby ground vibration due to traffic," Journal Structural Engineering Japan Society Civil Engineers **48A**, 437-444, (2002).

36. Z. Wang, A.C. Bovik, H.R. Sheikh and E.P. Simoncelli, "Image quality assessment from error visibility to structural similarity," IEEE Transactions on Image Processing. **13** (4): 600–61, (2004).
37. A. Pluen, P. A. Netti, R. K. Jain, and D. A. Berk, "Diffusion of macromolecules in agarose gels: comparison of linear and globular configurations," Biophysical Journal **77** (1), 542–552 (1999).
38. D. Magde, E.L. Elson and W.W. Webb, "Fluorescence correlation spectroscopy II. An experimental realization," Biopolymers **13** (1), 29-61 (1974).
39. R.R. Nair, H.A. Wu, P.N. Jayaram, I.V. Grigorieva and A.K. Geim, "Unimpeded permeation of water through helium-leak-tight graphene-based membranes," Science **335**, 442-444, (2012).
40. R.K. Joshi, P. Carbone, F.C. Wange, V.G. Kravets, Y. Su, I.V. Grigorieva, H.A. Wu, A.K. Geim and R.R. Nair, "Precise and ultrafast molecular sieving through graphene oxide membranes," Science **343**, 752-754 (2014).
41. C. Nethravathi, B. Viswanath, C. Shivakumara, N. Mahadevaiah and M. Rajamathi, "The production of smectite clay/graphene composites through delamination and co-stacking," Carbon **46**, 1773-1781 (2008).
42. R. Zhang, V. Alecrim, M. Hummelgard, B. Andres, S. Forsberg, M. Andersson and H. Olin, "Thermally reduced kaolin-graphene oxide nanocomposites for gas sensing," Scientific Reports **5**, 7676, (2015).
43. C. Nethravathi, J.T. Rajamathi, N. Ravishankar, C. Shivakumara and M. Rajamathi, "Graphite oxide-intercalated anionic clay and its decomposition to graphene-inorganic material nanocomposites," Langmuir **24**, 8240-8244 (2008).
44. J. Wang, C. Liu, Y. Shuai, X. Cui and L. Nie, "Controlled release of anticancer drug using graphene oxide as a drug-binding effector in konjac glucomannan/sodium alginate hydrogels," Colloids Surface B Biointerfaces **113**, 223-229 (2014).
45. G.W. Ashley, J. Henise, R. Reid and D.V. Santi, "Hydrogel drug delivery system with predictable and tunable drug release and degradation rates," Proceedings of the National Academy of Sciences USA **110**, 2318-2323 (2013).
46. E. Yablonovitch, "Inhibited spontaneous emission in solid-state physics," Physical Review Letters **58**(20), 2059-2062, (1987).
47. S. John, "Strong localization of photons in certain disordered dielectric structures," Physical Review Letters **58** (23), 2486- 2489, (1987).
48. J.W.S. Rayleigh, "On the remarkable phenomenon of crystalline reflexion described by Prof. Stokes," Philosophical Magazine **26**, 256-265, (1888).
49. V.P. Bykov, "Spontaneous Emission in a Periodic Structure," Soviet Journal of Experimental and Theoretical Physics **35**, 269-273, (1972).
50. K. Ohtaka, "Energy band of photons and low-energy photon diffraction," Physical Review B **19**(10), 5057-5067, (1979).
51. A. Colombi, P. Roux, S. Guenneau, P. Guéguen and R.V. Craster, "Forests as a natural seismic metamaterial: Rayleigh wave bandgaps induced by local resnances," Scientific reports **6**, 19238, (2016).
52. A. Colombi, D. Colquitt, P. Roux, S. Guenneau and R.V. Craster, "A seismic metamaterial: the resonant metawedge," Scientific Reports **6**, 27717, (2016).
53. P.A.M. Dirac, "The quantum theory of the electron," Proceedings of the Royal Society A **117**, 610, (1928).
54. K.S. Novoselov, A.K. Geim, S.V. Morozov, D. Jiang, M.I. Katsnelson, I.V. Grigorieva, S.V. Dubonos and A.A. Firsov, "Two-dimensional gas of massless Dirac fermions in graphene," Nature **438**(7065), 197-200, (2005).
55. S. Enoch, G. Tayeb, P. Sabouroux, N. Guerin, and P. Vincent, "A metamaterial for directive emission," Physical Review Letters **89**(21), 213902, (2002).
56. M. Silveirinha and N. Engheta, "Tunneling of electromagnetic energy through subwavelength channels and bends using ε-near-zero materials," Physical Review Letters **97**(15), 157403, (2006).
57. K. Sakoda, "Dirac cone in two- and three-dimensional metamaterials," Optics Express **20**(4), 3898-3917, (2012).
58. C. T. Chan, X. Huang, F. Liu, and Z. H. Hang, "Dirac dispersion and zero-index in two dimensional and three dimensional photonic and phononic systems," Progress In Electromagnetics Research B **44**, 163-190 (2012).

59. M. Kadic, T. Bückmann, R. Schittny, M. Wegener, "Experiments on cloaking in optics, thermodynamics and mechanics," Philosophical Transactions of the Royal Society A, DOI: 10.1098/rsta.2014.0357 (2015).
60. B. Gralak, G. Tayeb, S. Enoch, "Morpho butterflies wings color modeled with lamellar grating theory," Optics Express **9**(11), 567-578 (2001).
61. R.V. Craster, S. Guenneau, eds. Acoustic metamaterials (Springer Verlag, London, 2012).
62. S.H. Kim, M.P. Das, "Seismic waveguide of metamaterials," Modern Physics Letters B **26**, 1250105 (2012).
63. S. Krodel, N. Thome, C. Daraio, "Wide band-gap seismic metastructures," Extreme Mechanics Letters **4**, 111-117 (2015).
64. P.J. Lu, P.J. Steinhardt, "Decagonal and quasicrystalline Tillings in Medieval Islamic Architecture," Science **315**, 1106-1110 (2007).
65. B. Gralak, G. Arismendi, B. Avril, A. Diatta, and S. Guenneau, "Analysis in temporal regime of dispersive invisible structures designed from transformation optics," Physical Review B **93**, 121114(R) (2016).
66. S. Guenneau, "The physics of invisibility," https://blogs.royalsociety.org/publishing/the-physics-of-invisibility/
67. M. Miniaci, A. Krushynska, F. Bosia and N.M. Pugno, "Large scale mechanical metamaterials as seismic shields," New Journal of Physics **18**, 083041 (2016).
68. M. Moleron, S. Felix, V. Pagneux and E.O. Richoux, "Sound propagation in periodic urban areas," Journal Applied Physics **111**, 114906 (2012).
69. G. Finocchio, O. Casablanca, G. Ricciardi, U. Alibrandi, F. Garescì, M. Chiappini, and B. Azzerboni, "Seismic metamaterials based on isochronous mechanical oscillators," Applied Physics Letters **104**, 191903 (2014).
70. S. Brûlé, S. Enoch and S. Guenneau, "Flat seismic lens," arXiv:1602.04492v1
71. D. Bigoni, S. Guenneau, A.B. Movchan and M. Brun, "Elastic metamaterials with inertial locally resonant structures: Application to lensing and localization," Physical Review B **87**(17), 174303 (2013).
72. M. Farhat, S. Guenneau and H. Bağcı, "Exciting graphene surface plasmon polaritons through light and sound interplay," Physical Review Letters **111**(23), 237404 (2013).
73. P. Wang, F. Casadei, S. Shan, J.C. Weaver and K. Bertoldi, "Harnessing Buckling to Design Tunable Locally Resonant Acoustic Metamaterials," Physical Review Letters **113**, 014301 (2014).
74. M. Maldovan, "Sound and Heat Revolutions in Phononics," Nature **503**, 209-217 (2013).
75. R. Schittny, M. Kadic, S. Guenneau, and M. Wegener, "Experiments on Transformation Thermodynamics: Molding the Flow of Heat," Physical Review Letters **110**, 195901 (2013).
76. L. Domino, M. Tarpin, S. Patinet, and A. Eddi, "Faraday wave lattice as an elastic metamaterial," Physical Review E **93** 050202 (2016).
77. R. Schittny, M. Kadic, T. Buckmann, and M. Wegener, "Invisibility cloaking in a diffusive light scattering medium," Science **345** 427 (2014).
78. M. Farhat, P.Y. Chen, S. Guenneau, H. Bağcı, K.N. Salama and A. Alù, "Cloaking through cancellation of diffusive wave scattering," Proceedings Royal Society London A **472** (2192), 20160276 (2016).
79. W.H. Bragg and W.L. Bragg, "The Reflexion of X-rays by Crystals," Proceedings Royal Society London A **88** (605), 428-438 (1913).
80. H. Poincaré, "On the dynamics of the electron," Comptes rendus hebdomadaires des séances de l'Académie des sciences **140**, 1504-1508 (1905).
81. A. Einstein, "Zur elektrodynamik bewegter Körper," Annalen der Physik, **322** (10): 891-921 (1905).
82. A. Einstein, Relativity: The special and the general theory (Three Rivers Press, New York, 1961).
83. U. Leonhardt and T. G. Philbin, Geometry and Light: The Science of Invisibility (Dover Press, Mineola, 2010)
84. M. Farhat, P.Y. Chen, S. Guenneau, and S. Enoch, eds., "Transformation Wave Physics: Electromagnetics, Elastodynamics and Thermodynamics," (Pan Stanford Publishing, Singapore 2016)